\documentclass{article}

\usepackage{arxiv}

\usepackage[utf8]{inputenc} 
\usepackage[T1]{fontenc}    
\usepackage{hyperref}       
\usepackage{url}            
\usepackage{booktabs}       
\usepackage{amsfonts}       
\usepackage{nicefrac}       
\usepackage{microtype}      
\usepackage{lipsum}
\usepackage{graphicx}

\title{Coprocessors : failures and successes}

\date{July 25, 2019\\ English version of the paper presented in the French Conference COMPAS 2019.} 	

\author{
 Daniel Etiemble \\
Paris Sud University,\\
Computer Science Laboratory (LRI)\\
91405 Orsay - France\\
 \texttt{de@lri.fr}\\
}  

\begin{document}
\maketitle

\begin{abstract}
The appearance and disappearance of coprocessors by integration into the CPU, the success or failure of coprocessors are examined by summarizing their characteristics from the mainframes of the 1960s. The coprocessors most particularly reviewed are the IBM 360 and CDC-6600 I/O processors, the Intel 8087 math coprocessor, the Cell processor, the Intel Xeon Phi coprocessors, the GPUs, the FPGAs, and the coprocessors of manycores SW26010 and Pezy SC-2 used in high-ranked supercomputers in the TOP500 or Green500. The conditions for a coprocessor to be viable in the medium or long-term are defined.
\end{abstract}


\section{Introduction}
Since the early days of computers, coprocessors have been used to relieve the main processor (CPU) of certain “ancillary” tasks. These coprocessors have been or are being used for different tasks:
\begin{itemize}
\item I/O coprocessors
\item Floating-point coprocessors
\item Graphic coprocessors
\item Coprocessors for accelerating computation
\end{itemize}

The history of these coprocessors is diverse. Some have disappeared in the medium or short-term, following a technological breakthrough such as the invention of semiconductors, or the evolution of integrated circuit density. This is particularly the case of I/O coprocessors or floating-point coprocessors. Others like graphic coprocessors have a tumultuous history depending on their use for graphics or for high-performance computing or artificial intelligence. Coprocessors to accelerate computation appear and disappear like the Cell processor or Xeon Phi coprocessors. The best  supercomputers often use accelerators like GPUs or FPGAs.

As coprocessors are specific processors, they usually have their own instruction set and run their own programs. Transfers of programs and data between main processor and coprocessor must be managed. Having different execution models, the coexistence of different programming models is also an issue.

In this paper, through the review of several coprocessors, we examine the essential reasons that lead to the failure or success of a coprocessor.

\section{I/O coprocessors}
I/O coprocessors were used to offload the CPU from managing I/O  tasks in periods when the technology resources did not allow the CPU to perform them without significantly degrading performance on the program's main tasks. They were, therefore, used in mainframes and in the first generations of microprocessors.

\begin{figure}[htbp]
\centerline{\includegraphics  [width = 12 cm]{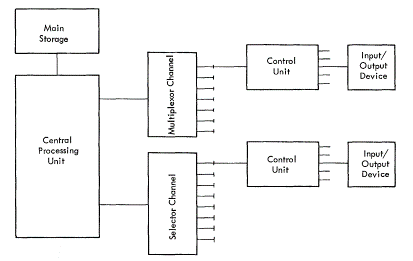}}
\caption{Processor and I/O coprocessors in the IBM 360}
\label{360}
\end{figure}

The IBM-360, manufactured in 1965, uses I/O processors \cite{ibm360} called channels that perform the transfers between the CPU and the I/O components (Figure \ref{360}). Multiplexer channels work at the byte level for slow devices. The selector channels are connected to fast devices.

The mainframe CDC6600 \cite {cdc6600}, also dating from 1965, has 10 peripheral processors. These peripheral processors (PPs) are connected to the I/O components through a switching circuit. An interesting feature of these PPs is that it is the first implementation of fine-grain hardware multithreading. At each clock cycle of the PPs, one goes from the execution of an instruction by a PP to the execution of an instruction by the following PP. This technique allowed each PE to access the main memory every ten cycles (latency of this memory).

For 8-bit and 16-bit microprocessors, the low integration density led to I/O tasks being moved to a coprocessor named Intel 8089. The Intel 8037 DMA controller also assisted the 8086 CPU.

For Intel microprocessors, the management of I/Os with controlled DMA and interrupts is ensured by the CPU starting with the 80186 CPU.

\section{Floating-point coprocessors}

Floating-point operators are far more complex than integer operators. The first generations of microprocessors had to use a math coprocessor.
The typical example is the Intel 8087 coprocessor.
 Like Motorola's 68881 and 68882 coprocessors from the same era, the 8087 is working on an 80-bit floating-point format that can handle single and double-precision formats.

The 8087 highlights a peculiarity that has had significant consequences on the suite of Intel implementations of floating-point computing in subsequent generations of microprocessors. It uses a stack of eight 80-bit  registers. Using the instruction set classification defined by Hennessy and Patterson (\cite {henpat}), the instructions use a slightly modified variant of the stack mode (0 operand and 0 memory access per instruction) while the x86 instructions use the mode (2,1): two operands per instruction, one of which may be a memory operand.

Versions 80187, 80287 and 80387 have been produced. When the integration density allowed it (CPU 80486 in 1989), the floating-point operators of the 8087 were integrated into the CPU chip. Since that date, for the upward binary compatibility, all Intel CPUs integrate the 8087 operators and the corresponding execution mode. It turns out that the floating-point performance of Intel processors have suffered from the 8087 execution mode. Intel has worked around the problem by defining a second set of floating-point instructions. The SIMD instructions, especially from SSE, use SIMD registers, called XMMs for SSE. In addition to the SIMD floating instructions defined for single and double precision, Intel has defined "scalar" versions of the SIMD instructions, which only work on one element of the vector. The SIMD scalar instructions have a format (2.1) like the other instructions. It should be noted that this new set of floating instructions is not a substitute, but is added to the x87 instructions.
 
The increase of the integration density has thus made it possible to integrate the math coprocessors into the processor chip. However this integration can raise issues if the instruction formats are different.

\section{Graphic coprocessors}
Graphics cards are a typical example of coprocessors used for graphic rendering. They have evolved to fully programmable graphics processors (GPUs) also used as computing accelerators.

\section{Coprocessors to speed-up computationsl}
Computation accelerators are typical examples of coprocessors. These coprocessors have very different stories, as we will see with the examples of the Cell processor, Xeon Phi coprocessors, and GPUs.

\subsection{Cell processor}
\begin{figure}[htbp]
\centerline{\includegraphics  [width = 12 cm]{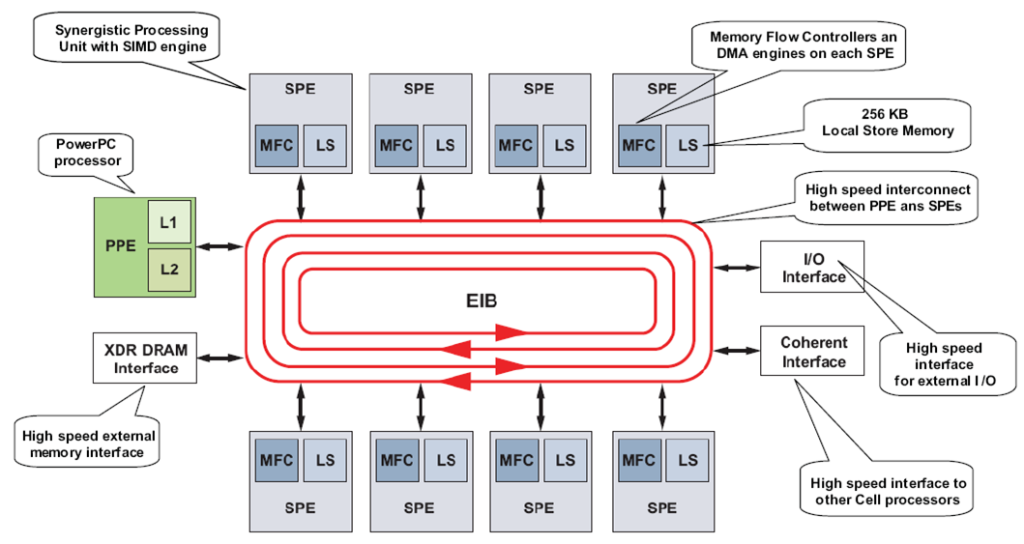}}
\caption{Cell processor}
\label{Cell}
\end{figure}

The Cell \cite {cell} is a processor designed jointly by IBM, Sony and Toshiba, introduced in February 2005. Originally intended for video games (Sony's PlayStation 3), its floating-point performance has made it used in areas such as high performance computing, image processing, machine vision, etc.
The architecture of the Cell is shown in Figure \ref {Cell}. Next to a PowerPC processor called PPE, there are eight coprocessors called SPE. An SPE consists of:

\begin{itemize}
\item a 2-way in-order superscalar processor executing SIMD 128-bit single and double-precision instructions.
\item a 256-KB local memory.
\item a memory controller with DMA transfers.
\end{itemize}
Code and data transfers between the processor and the SPE are controlled by software.
Although the Cell processor has had some success, it has been relatively short-lived. IBM abandoned the Cell in late 2009, and its limits appeared very quickly for high-end computing.

The short live of the Cell processor can be explained by several considerations: :
\begin{itemize}
\item Each SPE coprocessor does not have a computation capacity very different from that of the PPE, which also has 128-bit SIMD instructions. It is the number of SPEs that increases the performance. In addition to the communication problems between the main processor and the eight SPEs, there are all the software problems associated with a heterogeneous parallel architecture.
\item Transfers between PPE and the different SPEs with their local memory are done by DMA under software control. If this approach is found in 2018 in successful processors with several dozens of cores, it is much easier to use a multi-core with an equivalent number of cores with parallel programming in shared memory based on cache coherency. In 2010, when IBM retired, a 10-core (and 20-thread) Xeon CPU already existed.
\end{itemize}

A first rule appears: there must be at least an order of magnitude between the performance of a coprocessor compared to  the main processor performance for the coprocessor being viable in the medium term.

\subsection{Xeon Phi coprocessors}
Xeon Phi are coprocessors developed by Intel for high performance computing. There were two generations of Xeon Phi whose code names were KNC (Knights Corner) and KNL (Knights Landing). KNL also had an independent processor version.
The Intel Xeon Phi coprocessor is an extension of multi-core features, including:
\begin{itemize}
\item Software compatibility with the Intel x86 instruction set
\item Memory hierarchy with cache coherency: logically shared memory..
\item Virtual memory with page translation and TLB
\item Ring or 2D grid interconnection network as in multi-cores
\end{itemize}

\begin{figure}[htbp]
\centerline{\includegraphics  [width = 10 cm]{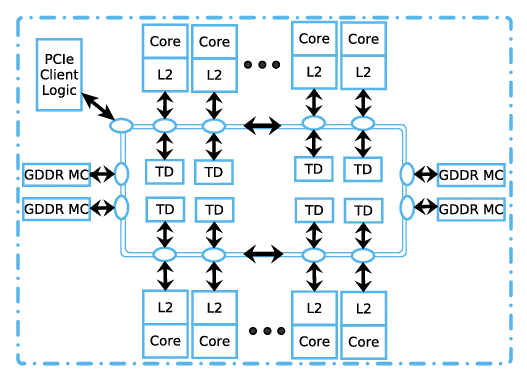}}
\caption{ KNC Xeon Phi}
\label{KNC}
\end{figure}

The KNC generation \cite {KNC}  was introduced in 2012. The  architecture is shown in Figure \ref {KNC}. Up to 61 cores can be connected to a bidirectional ring. Each core is a 2-way in-order superscalar processor derived from the Pentium P54C architecture, with simultaneous multithreading of 4 threads [3]. It has a SIMD unit called a Vector Processing Unit (VPU) with 512 bit vectors and can therefore execute 16 operations in single precision or 8 operations in double precision per cycle. 8 GB of GDDR5 memory are also connected to the bus. The ring is connected to the host processor through a PCIe link.

\begin{figure}[htbp]
\centerline{\includegraphics  [width = 11 cm]{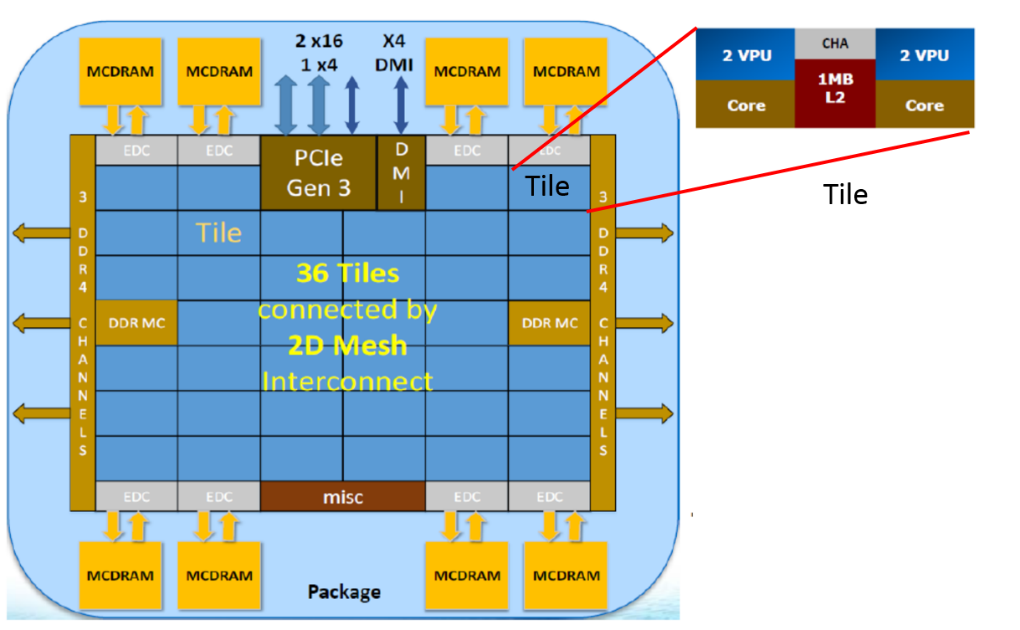}}
\caption{KNL Xeon Phi}
\label{KNL}
\end{figure}

The KNL architecture \cite {KNL}, available from 2016, retains the same architectural concepts of compatibility with the entire Intel range. It is presented in Figure \ref {KNL}. Compared to KNC, it introduces several modifications, at different levels:

\begin{itemize}
\item The ring  network is replaced by a 2D grid, operating with three different modes called respectively "All to All", "Quadrant" and "Sub-Numa clustering". This evolution is the same as that observed for Intel's multi-cores, from the rings of the Nehalem and Broadwell architectures to the 2D grid of Skylake and following architectures.
\item The 2D grid interconnects 36 tiles consisting of two cores, two SIMD (VPU) units per core sharing a 1 MB L2 cache.

\item Each core is a 2-way out-of-order superscalar processor derived from the Silvermont architecture used in Atom, Celeron, and Pentium low-power processors. The core uses simultaneous multithreading with 4 threads. VPUs implement the 512-bit AVX version.
\item Two types of memory are used: MCDRAM (Multi-Channel DRAM), which is a very high speed 3D DRAM and DDR4, which is the classic DRAM memory. There are three types of operation: 1) cache mode, where the MCDRAM serves as a level 3 cache, 2) the "flat" mode where MCDRAM and DDR4 constitute the memory and the coherence is managed by software, and 3) the hybrid mode where the MCDRAM serves partly as cache and partly as main memory.

\end{itemize}

Xeon Phi coprocessors have met some success. By November 2018, they were used by 18 of the 500 TOP 500 supercomputers. But in the same list, there were 128 supercomputers using NVidia GPUs. The Xeon Phi versions have not been able to compete efficiently with GPUs, which led to the discontinuation of the Knight Hills project (third generation of Xeon Phi) and the announced shutdown of KNL production in mid-2018 .

\subsection{GPUs}
Graphics processors became fully programmable in the mid-2000s, with AMD's Xenos GPUs and NVidia's GeForce6600. From graphic processors, they have also become essential for the high-end computing, as shown by their presence as accelerators in a large number of TOP 500 supercomputers.
The block diagram of a GPU (Fermi architecture {\cite {Fermi}) is given in Figure \ref {GPU}. The architecture is decomposed into three levels: the GPU comprises several multiprocessors, each multiprocessor having a lot of CUDA cores. We do not detail all the elements of the different parts of this diagram.

\begin{figure}[htbp]
\centerline{\includegraphics  [width = 16 cm]{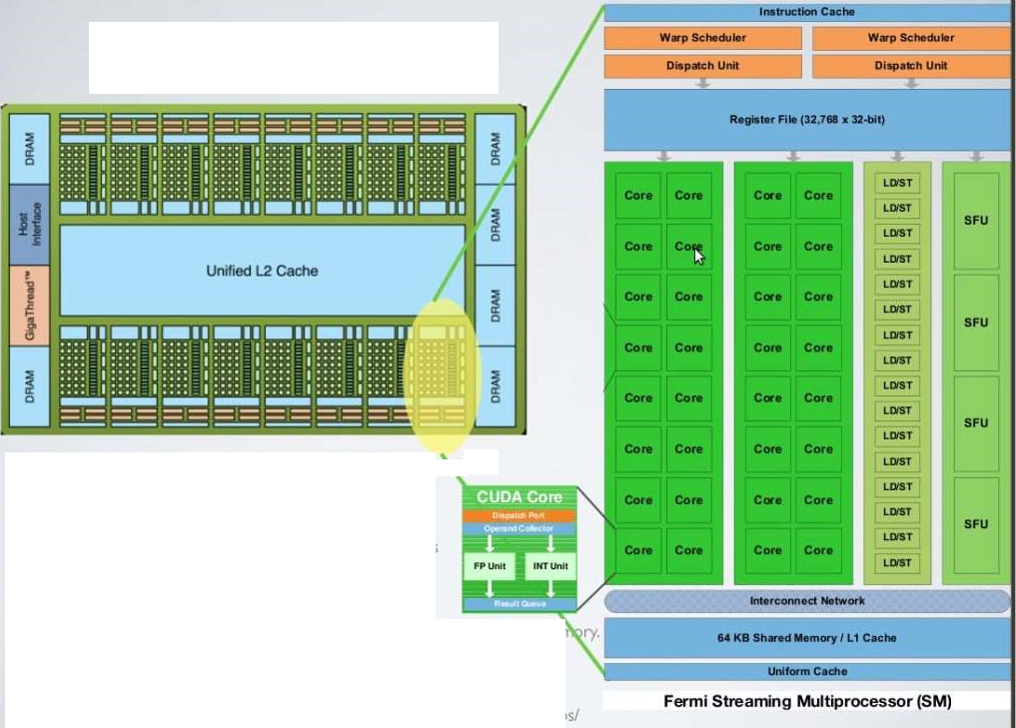}}
\caption{NVidia Fermi GPU architecture}
\label{GPU}
\end{figure}

The key point that explains the success of GPUs for high-end computing is that the cores of a GPU (CUDA cores in Figure \ref {GPU}) are totally different from CPU cores as shown in Figure \ref {GPUDif} . The difference are:
\begin{itemize}
\item The CPU is designed for complex applications, with a significant control logic, a small number of operators (even extended with SIMD extensions), a hierarchy of large caches (L1, L2, L3 ...), instructions with a low latency, etc. 
\item The GPU is designed for massive parallelism with a very large number of computing cores. There are many computations by memory access. Pipelines have a very large number of stages (hundreds). Latencies are important. Execution bandwith is very high.
\end{itemize}

\begin{figure}[htbp]
\centerline{\includegraphics  [width =12 cm]{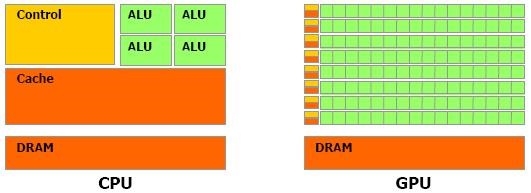}}
\caption{CPU and GPU differences}
\label{GPUDif}
\end{figure}

This difference is reflected in the maximum performance level, as shown in Figure \ref {GPUPerf}.
GPUs have different execution and programming models from those of CPUs and are used as coprocessors. The CPU-GPU connection is made via  PCIe or NVlink type connection if the two components  have their own memory, or via a common memory, as shown in Figure \ref {GPULinks}.
A disadvantage of high performance GPUs is the energy consumption, which reaches several hundred Watts.

\begin{figure}[htbp]
\centerline{\includegraphics  [width = 12 cm]{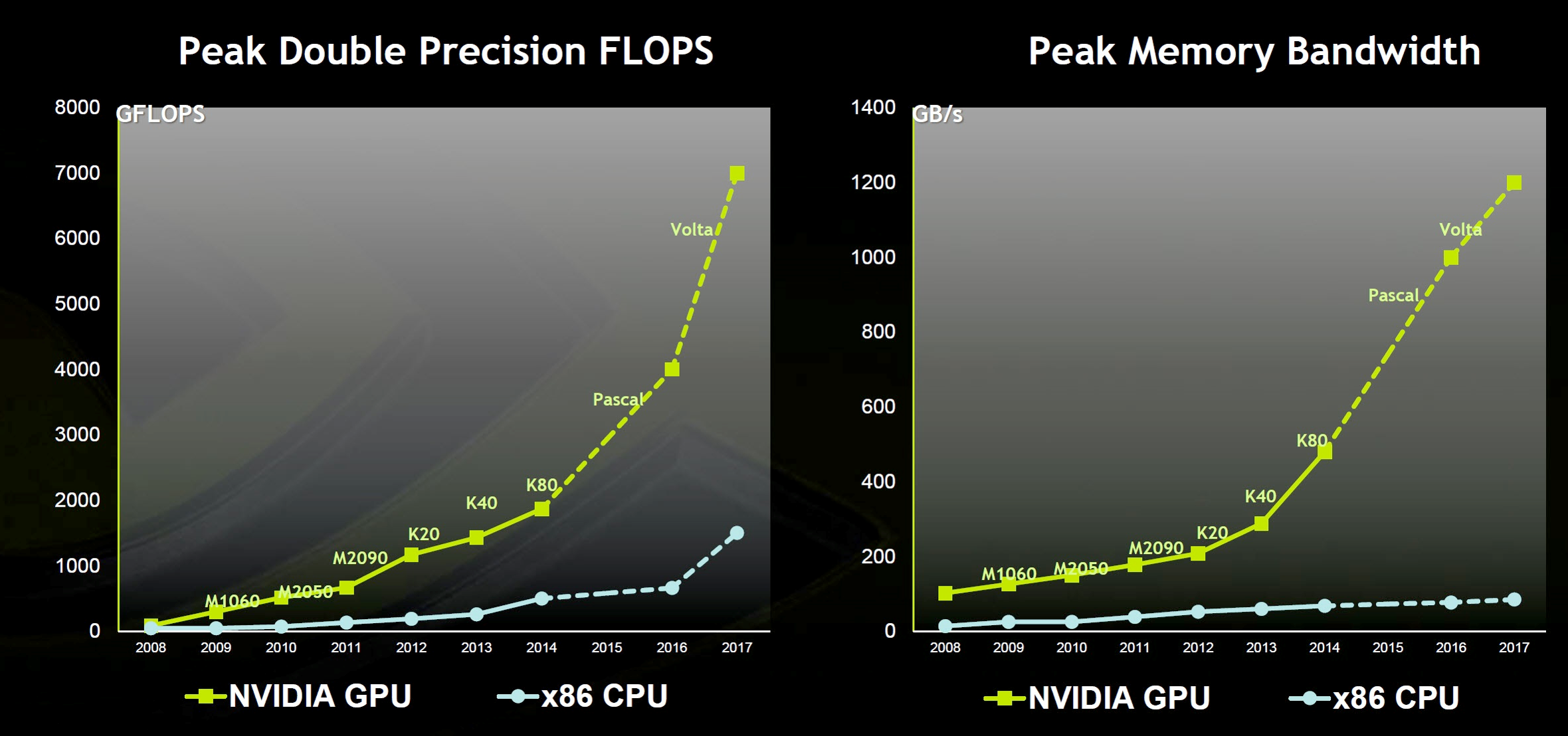}}
\caption{Comparing GPU Nvidia and CPU x86 performance}
\label{GPUPerf}
\end{figure}

\begin{figure}[htbp]
\centerline{\includegraphics  [width = 10 cm]{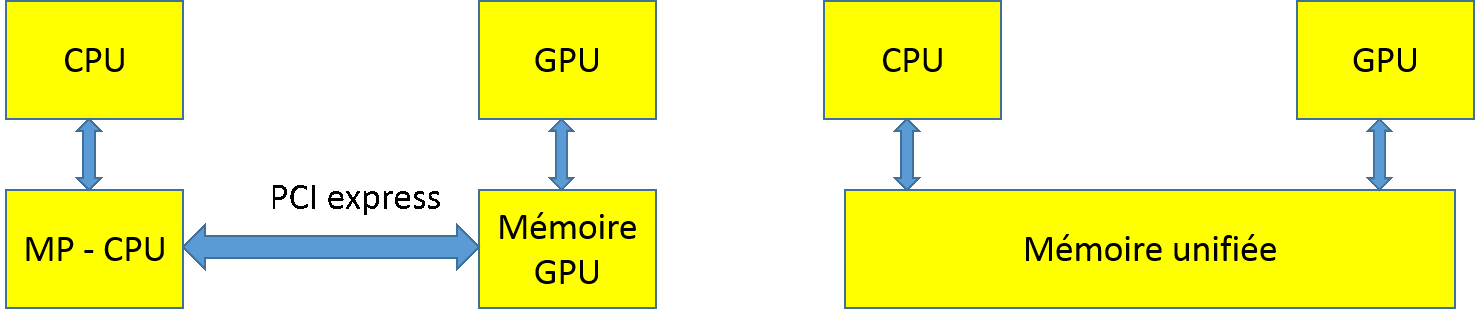}}
\caption { CPU-GPU interconnect}
\label{GPULinks}
\end{figure}

We can deduce some conclusions about the reasons for the current success of GPUs:
\begin{itemize}
\item The coprocessor should not be a simple processor aid, but must provide a significant performance gain, at least an order of magnitude, compared to the most optimized versions of programs running on the processor.
\item The overheads associated with transfers between CPUs and coprocessors must be minimized, and the interconnect shift from PCIe to unified memory is an example of this minimization for GPUs.
\item The coexistence between two different software environments must be ensured efficiently. Successive versions of CUDA for NVidia GPUs and OpenCL (with less success) have simplified the software developments.
\end{itemize}

\subsection{FPGAs}
As GPUs have evolved from simple graphics processors into high-end computing accelerators, FPGA chips have evolved from components to implement combinational and sequential logic into true Systems-on-Chip (SoCs) while retaining their primary interest: hardware programming. In addition to the original logic blocks, high-end FPGAs have memory blocks, signal processing blocks (DSPs), "hard" processors, and more.
There are two types of problems associated with FPGAs:
\begin{itemize}
\item Clock frequencies have long been an order of magnitude lower than those of CPUs of the same era. This gap is closing: the Intel Stratix 10 FPGAs have a maximum clock frequency of 1 GHz.
\item While programming of FPGA hardware configurations has long been done using low-level specific languages such as VHDL or Verilog, higher-level approaches are now available (OpenCL, hardware compilers) allowing software developers to use FPGAs as accelerators without having to take into account the material developments.
\end{itemize}
FPGAs can naturally benefit from the use of particular arithmetic formats for applications such as deep neural networks. They have the advantage of lower consumption compared to GPUs.
While FPGAs are not yet used as coprocessors in TOP 500 supercomputers, there is growing interest in using them in cloud hardware platforms (data centers). FPGAs can be considered as being in the GPU situation in the early 2000s, in the early days of their use in high-end computing.
\subsection{Manycores}
The manycores processors used in the most powerful supercomputers generally consist of a control processor and a large number of compute processors that act as coprocessors of the control processor. For example, this is the case for the following processors:
\begin{itemize}
\item The SW26010 \cite {SW26010} used in the Sunway TaihuLight supercomputer consists of a 64-bit control RISC (MPE) and a cluster of 64 computing cores (CPE).
\item The Pezy SC-2 used in the Shoubu System2 supercomputer has a multi-core MIPS P6600 with 6 cores for control and 2048 cores organized in 3 levels of clusters for calculations.
\end{itemize}
While the previous manycores consume several hundred watts, manycores consuming only tens of watts like Kalray's  MPPAs\cite {Kalray} are used as coprocessors in acceleration cards. 

\section{Concluding remarks}
The fate of the coprocessors depends on their nature and the type of connection with the processor, both from a hardware and software point of view.
When the tasks performed by the coprocessor were directly compatible with the tasks of the processor, such as I/O management or floating-point computing, it was the lack of resources available in the processor that justified the existence of a coprocessor. As soon as the CPU resources have grown sufficiently, the coprocessor operations have been directly integrated into the processor. This was the case for I/O coprocessors and floating-point coprocessors.
Successful coprocessors have three main characteristics:
\begin{enumerate}
\item They have an execution model completely different from the CPU one. This justifies their own existence. It is difficult to directly integrate them in the CPU.
\item They have a sufficient performance gain compared to the CPU: at least an order of magnitude.
\item  They have a mature software environment to program the CPU + coprocessor set in a relatively simple and efficient way. This does not exclude that specific optimizations are still needed to get the best performance.
\end{enumerate}
GPUs meet all three criteria. For FPGAs, the software environments (3) are still insufficient. In contrast, none of the three criteria was satisfied for the Cell processor. The processor + SIMD coprocessor combination did not perform much better than conventional multi-cores with SIMD extensions, and parallel programming with distributed memory was far more complex than shared memory and cache hierarchy. Similarly, condition 2 was not really fulfilled for Xeon Phi coprocessors, even though they were used in supercomputers.

\bibliographystyle{unsrt}  


\begin{thebibliography}{1}

\bibitem{ibm360} IBM Systems Reference Library, \textit{IBM System/360 Principles of Operation}, http://bitsavers.trailing-edge.com/pdf/ibm/360/princOps/$A22-6821-0\_360$PrincOps.pdf
 
\bibitem{cdc6600} Control Data, \textit{Control Data 6400/6500/6600 Computer Systems Reference Manual},  http://www.textfiles.com/bitsavers/pdf/cdc/$6x00/60100000D\_6600$refMan\_Feb67.pdf

\bibitem{henpat} Patterson (D.A.) and Hennessy (J.L.), \textit{Computer Architecture, A Quantitative Approach}, $2^{nd}$ edition, Morgan Kaufmann, 1996

\bibitem{cell} Gschwind (M.), Hofstee (H.P.), Flachs (B.), Hopkins (M.), Wanatabe (Y.) and Yamazaki [T.), \textit{Synergistic Processing in Cell’s Multicore Architecture}, IEEE Micro, $26(2), 10-24, doi: 10.1109\/mm.2006.41$

\bibitem{KNC} Rahman (R), \textit{Intel® Xeon Phi™ Coprocessor Architecture and Tools: The Guide for Application Developers}, https://www.oreilly.com/library/view/intel-xeon-phitm/$9781430259268$/

\bibitem{KNL} Sodani (A), \textit {Knights Landing (KNL): 2nd Generation Intel® Xeon Phi™ Processor}, https://www.alcf.anl.gov/files/HC27.25.710-Knights-Landing-Sodani-Intel.pdf  

\bibitem{Fermi} NVidia, \textit{ NVIDIA’s Next Generation CUDA Compute Architecture: Fermi}, http://www.nvidia.com/content/PDF/fermi\_white\_papers/\\NVIDIA\_Fermi\_Compute\_Architecture\_Whitepaper.pdf 

\bibitem{SW26010} Xu (Z), Lin (J), Matsukoa (S), \textit {Benchmarking SW26010 Many-core Processor}, Proceedings 2017 IEEE International Parallel and Distributed Processing Symposium Workshops, pp 743-752, https://hpc.sjtu.edu.cn/IPDPSW2017\_Benchmarking.pdf  

\bibitem{Kalray} Kalray, \textit{ Deep Learning for High Performance Embedded Applications}, White paper, 2017, http://www.eenewseurope.com/Learning-center/kalray-deep-learning-high-performance-applications.pdf   


\end{thebibliography}

\end{document}